\documentclass[prb,
amsmath,
amssymb,
noeprint,
superscriptaddress
]{revtex4-2}

\usepackage{graphicx}
\usepackage{bm}% bold math
\usepackage{hyperref}
\usepackage[OT2,T1]{fontenc}
\DeclareSymbolFont{cyrletters}{OT2}{wncyr}{m}{n}
\DeclareMathSymbol{\Sha}{\mathalpha}{cyrletters}{"58}
\newcommand{\wn}{$\textrm{cm}^{-1}$}

\begin{document}
%%%%%%%%%%%%%%%%%%%%%%%%%%  title  %%%%%%%%%%%%%%%%%%%%%%%%%%
\title{Taking advantage of multiplet structure for lineshape analysis in Fourier space}

%%%%%%%%%%%%%%%%%%%%%%%%%%  addresses  %%%%%%%%%%%%%%%%%%%%%%%%%%
\author{A. Beckert}
\affiliation{Laboratory for Micro and Nanotechnology, Paul Scherrer Institut, CH-5232 Villigen PSI, Switzerland.}%
\affiliation{Laboratory for Solid State Physics, ETH Zurich, CH-8093 Zurich, Switzerland.}
\author{H. Sigg}
\affiliation{Laboratory for Micro and Nanotechnology, Paul Scherrer Institut, CH-5232 Villigen PSI, Switzerland.}

\author{G. Aeppli}
\affiliation{Laboratory for Micro and Nanotechnology, Paul Scherrer Institut, CH-5232 Villigen PSI, Switzerland.}
 \affiliation{Laboratory for Solid State Physics, ETH Zurich, CH-8093 Zurich, Switzerland.}
 \affiliation{Institute of Physics, EPF Lausanne, CH-1015 Lausanne, Switzerland.}

% \author{Adrian Beckert,\authormark{1,2,4} Hans Sigg,\authormark{1} and Gabriel Aeppli\authormark{1,2,3,5}}

% \address{
% \authormark{1}Laboratory for Micro and Nanotechnology, Paul Scherrer Institut, CH-5232 Villigen PSI, Switzerland.\\
% \authormark{2}Laboratory for Solid State Physics, ETH Zurich, CH-8093 Zurich, Switzerland.\\
% \authormark{3}Institute of Physics, EPF Lausanne, CH-1015 Lausanne, Switzerland.
% }
% \email{\authormark{4}adrian.beckert@psi.ch}
% \email{\authormark{5}gabriel.aeppli@psi.ch}

%%%%%%%%%%%%%%%%%%%%%%%%%%  abstract  %%%%%%%%%%%%%%%%%%%%%%%%%%
\begin{abstract}
Lineshape analysis is a recurrent and often computationally intensive task in optics, even more so for multiple peaks  in the presence of noise. We demonstrate an algorithm which takes advantage of peak multiplicity ($N$) to  retrieve line shape information. The method is exemplified via analysis of Lorentzian and Gaussian contributions to individual lineshapes for a practical spectroscopic measurement and benefits from a linear increase in sensitivity with the number $N$.  The robustness of the method and its benefits in terms of noise reduction and order of magnitude improvement in run-time performance are discussed.
\end{abstract}

%%%%%%%%%%%%%%%%%%%%%%%%%%  title  %%%%%%%%%%%%%%%%%%%%%%%%%%
\maketitle

%%%%%%%%%%%%%%%%%%%%%%%%%%  body  %%%%%%%%%%%%%%%%%%%%%%%%%%
\section{Introduction}
\label{intro}
Analysis of signal lineshapes is a prominent problem and a theme of importance in  physics, chemistry and biomedicine. Ranging from spectroscopy \cite{vandehulst1947,galatry1961,rautian1967} to scattering techniques \cite{stokes1948,wilson1962,keijser1982,maletta1982}, the lineshape can reveal underlying physical processes. For example, relaxation dynamics very commonly give rise to exponential decays in time which correspond in spectroscopy to Lorentzian peaks in frequency, while static disorder and instrumental effects typically induce Gaussian distributions of characteristic frequencies. Together, the two phenomena yield signal lineshapes which are convolutions of Lorentzians with Gaussians, objects referred to as Voigt lineshapes. The accurate parametrization of a Voigt lineshape retrieves the Lorentzian contribution  which \textit{e.g.} quantifies correlation lengths in X-ray and  neutron scattering \cite{maletta1982} 
and coherence times of quantum systems in frequency-dependent spectroscopy \cite{dirac1927,schober1933,orear1967}; and the Gaussian part which is due to extrinsic factors such as grain size distribution  in x-ray diffraction \cite{keijser1982,enzo1988} and to spatial inhomogeneity in optical spectroscopy \cite{galatry1961,hutchinson2002}.\\
Approaches to lineshape determination deal with efficient numeric approximations \cite{mclean1994,liu2001} or tackle the deconvolution of single peaks in Fourier space \cite{shin2014}. However, the calculation and fitting of a Voigt profile remains computationally intensive. In case of finite-periodic signals, frequently encountered \textit{e.g.} for electronic multiplets of ions in solids \cite{matmon2016} or rotational spectra of molecules \cite{planck1917,jung1931,albert2011}, the problem is exacerbated by the need to sum over Voigt profiles.\\
In this article we extend previous work on single Voigt analysis (see \cite{shin2014} for an overview) and exploit the regular spacing for a robust and direct determination of the individual lineshape in Fourier space. We fit an exponential to the envelope of the Fourier transform of the signal, which is parametrized only by the Lorentzian and Gaussian contribution to the individual line width. The method quantifies these contributions without the need for involved multi-Voigt profile analysis. Results are also readily checked by direct visual inspection of the Fourier transform. Furthermore, our method is computationally less expensive and it's sensitivity increases linearly in $N$, the number of profiles. Although we focus on the common case of Voigt-shaped profiles, we expect our method will prove advantageous for other line-profiles, such as listed in refrence~\cite{ruland1968}, as well.\\
We first derive the mathematical foundations of the method in section~\ref{method}. In section~\ref{application} we apply the procedure to typical experimental data from solid-state spectroscopy, in this case a rare-earth doped crystal. Finally, in section~\ref{robust} we discuss the method's performance when the initial assumption of regular spacing is relaxed.

\section{Derivation of the Method}
\label{method}
We consider a real, finite-periodic signal $S({{x}})=\sum_iS_i(x)$, consisting of $N$ profiles $S_i(x)$ spaced by $\Delta{{x}}$. We focus on the common cases of individual signal profiles $S_i(x)$ of Lorentzian, Gaussian and Voigt shapes. The general case for other classes of lineshapes is addressed subsequently.

\subsection{Dirac delta model}
\begin{figure}[htbp]
\centering\includegraphics[width=7cm]{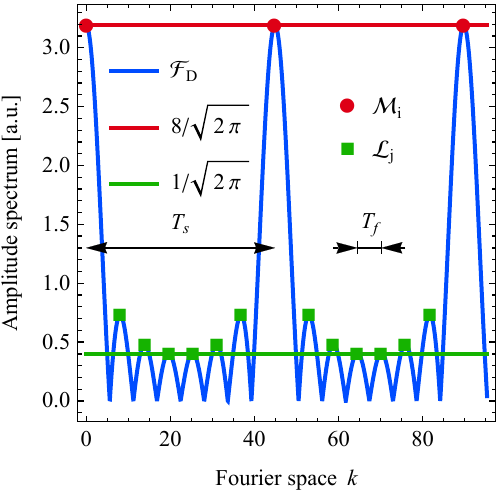}
\caption{\noindent Amplitude spectrum $|\mathcal{F}_\mathrm{D}|$ of signal $S_\mathrm{D}(x)$ for $N=8$ and $A=1$ (blue). The maxima $\mathcal{M}_i$ for $i\in\{1,2,3\}$ are depicted in red and the inferior local maxima $\mathcal{L}_j$ in green. The uniformity of the amplitude of the $\mathcal{M}_i$ is highlighted by the constant red line at $8/\sqrt{2\pi}$, with spacing $T_s$ and the minimal amplitude $\mathcal{L}_{j,\mathrm{min}}$ by the constant green line at $\sim1/\sqrt{2\pi}$, spaced by $T_f$. The limits of $\mathcal{M}_i$ and $\mathcal{L}_{j,\mathrm{min}}$ are derived from equation~\ref{diracFT}.}
\label{fig1}
\end{figure}

For simplicity we first model $S({{x}})$ as a series of $N$ Dirac delta functions $\delta({{x}})$, with $N$ even, symmetrically disposed around ${{x}}=0$. Definitions and a detailed derivation are  in section~\ref{defs} and~\ref{detder} of the Appendix. The Fourier transform $\mathcal{F}:{{x}}\rightarrow k$ applied to the Dirac model $S_\mathrm{D}({{x}})$ is given by
\begin{equation}
\label{diracFT}
\begin{split}
\mathcal{F}_\mathrm{D}=\mathcal{F}[S_\mathrm{D}({{x}})]=\frac{A}{\sqrt{2\pi}}\frac{\mathrm{sin}(N\Delta x k/2)}{\mathrm{sin}(\Delta x k/2)}
\end{split}
\end{equation}
where $A$ represents the integrated area of an individual signal profile. Figure~\ref{fig1} shows  the amplitude spectrum $|\mathcal{F}_\mathrm{D}|$ ,
which exhibits a long periodicity $T_s=2\pi/\Delta{{x}}$ modulated by a short periodicity $T_f=T_s/N$. Note the frequency doubling of $|\mathcal{F}_\mathrm{D}|$ w.r.t $\mathcal{F}_\mathrm{D}$. The occurrence of $T_s$ in $|\mathcal{F}_\mathrm{D}|$ stems from the regular spacing of the $S_i(x)$, and $T_f$ from the length of the multiplet signal, which can be understood as the product of a rectangular window function with a Dirac comb (details in Appendix~\ref{altder}) in analogy to the result of an $N-$periodic diffraction grating \cite{klein1986}. We denote the $i$-th maximum of $|\mathcal{F}_\mathrm{D}|$ as $\mathcal{M}_{i}$, \textit{i.e.} $|\mathcal{F}_\mathrm{D}(i\times T_s)|=\mathcal{M}_{i}~\forall i\in\mathbb{N}_0$. The inferior local maxima are labeled by $\mathcal{L}_j$. From equation~\ref{diracFT} we obtain $\mathcal{M}_i\propto N$ for $\Delta{{x}k/2}~\mathrm{mod}~\pi=0$ with the rule of Bernoulli-de l'H\^{o}spital and $\mathcal{L}_{j,\mathrm{min}}= 1/\sqrt{2\pi}$ does not depend on the multiplet number $N$. $\mathcal{M}_i\propto N$ represents the scaling of the sensitivity of our Fourier lineshape analysis (FLA) method.

\subsection{Lorentzian and Gaussian lineshapes}
\begin{figure}[htbp]
\centering\includegraphics[width=8.56cm]{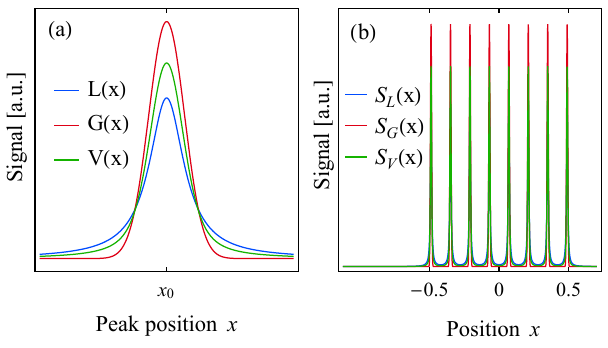}
\caption{\noindent Signal $S(x)$ lineshape illustrations. (a) Lorentzian $L(x)$ (blue), Gaussian $G(x)$ (red) and Voigt $V(x)$ (green) lineshape with identical center $x_0$, area and FWHM. (b) Sum of eight equally spaced Lorentzian $L(x)$, Gaussian $G(x)$ and Voigt $V(x)$ profiles, each with identical area and FWHM.}
\label{fig2}
\end{figure}

\begin{figure}[htbp]
\centering\includegraphics[width=7cm]{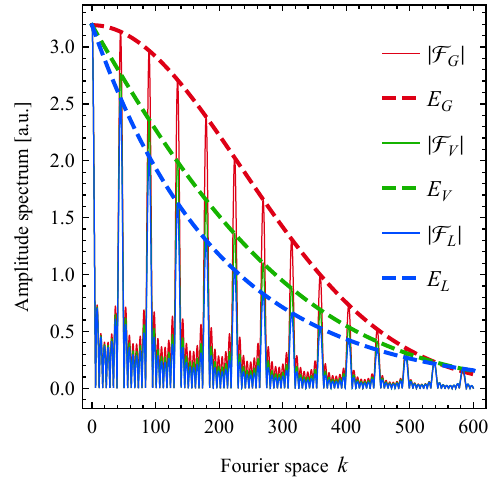}
\caption{\noindent Fourier transform of the signal functions shown in figure~\ref{fig2}(b). $|\mathcal{F}_\mathrm{L}|$ (blue), $|\mathcal{F}_\mathrm{G}|$ (red) and $|\mathcal{F}_\mathrm{V}|$ (green). The first few $\mathcal{M}_{k,i}$ are depicted for all three cases, together with the envelopes (dashed, corresponding color) defining their amplitudes.}
\label{fig3}
\end{figure}

We continue for Lorentzian-shaped signal lines $S_i(x)=L({{x}})$, as illustrated in figure~\ref{fig2}(a) and \ref{fig2}(b) for $N=8$.
The shape is given by
\begin{equation}
\label{lorentz}
L({{x}},{{x}}_0,\Gamma,A)=\frac{A}{2\pi}\frac{\Gamma}{({{x}}-{{x}}_0)^2+(\frac{1}{2}\Gamma)^2}
\end{equation}
with full-width-at-half-maximum (FWHM) $\Gamma$, center peak coordinate ${{x}}_0$ and integrated area $A=\int_{-\infty}^\infty L({{x}})\mathrm{d}{{x}}$. An $N$-fold repeated signal with spacing $\Delta{{x}}$ is represented by
\begin{equation}
S_\mathrm{L}({{x}})=\sum_{n=({1-N})/{2}}^{(N-1)/2}L({{x}},x_0+n\Delta{{x}},\Gamma,A)
\end{equation}
and its Fourier transform $\mathcal{F}_\mathrm{L}=\mathcal{F}[S_\mathrm{L}({{x}})]$ as
\begin{equation}
\label{lorentzFT}
\mathcal{F}_\mathrm{L}=\mathrm{exp}\left(-\frac{1}{2}\Gamma k\right)\times\frac{A}{\sqrt{2\pi}}\frac{\mathrm{sin}(N\Delta x k/2)}{\mathrm{sin}(\Delta x k/2)}
\end{equation}
for $x_0=0$. We see that $\mathcal{F}_\mathrm{L}$ (eq.~\ref{lorentzFT}) differs from $\mathcal{F}_\mathrm{D}$ (eq.~\ref{diracFT}) but by an exponentially decaying prefactor
\begin{equation}
\mathcal{F}_\mathrm{L}=\mathrm{exp}\left(-\frac{1}{2}\Gamma k\right)\times\mathcal{F}_\mathrm{D}.
\end{equation}
Figure~\ref{fig3} visualizes this result for $N=8$.
The zero-order maximum $\mathcal{M}_{k,0}$ exhibits the same amplitude as for equation~\ref{diracFT} but the amplitudes $\mathcal{M}_{i}$ for $i>0$ decay with $E_\mathrm{L}(k)=\mathrm{exp}({-\frac{1}{2}\Gamma k})$ and the sensitivity $\mathcal{M}_{i}\propto N$ of $\mathcal{F}_\mathrm{D}$ still holds. In the same way we obtain for a Gaussian model $S_G(x)$
\begin{equation}
\label{gaussFT}
\mathcal{F}_\mathrm{G}=\mathcal{F}\left[S_\mathrm{G}({{x}})\right]=\mathrm{exp}\left(-\frac{1}{2}\sigma^2 k^2\right)\times\mathcal{F}_\mathrm{D}
\end{equation}
which again exhibits the same periodicity $T_s$ as equation~\ref{diracFT} and equation~\ref{lorentzFT}. The exponential prefactor defining the envelope $E_\mathrm{G}(k)$ of $\mathcal{F}_\mathrm{G}$ is now given by $E_\mathrm{G}(k)=\mathrm{exp}({-\frac{1}{2}\sigma^2 k^2})$. The amplitude reduction of $\mathcal{M}_{i}$ for $i>0$ thus depends on $\sigma$ which is related to the Gaussian FWHM
\begin{equation}
\label{fwhmg}
f_\mathrm{G}=2\sqrt{2\mathrm{ln}(2)}\sigma.
\end{equation}

\subsection{Voigt profile model}
The Voigt lineshape $V({{x}})$ is defined as the convolution $(*)$ of a Lorentzian with a Gaussian function:
\begin{equation}
V({{x}})=L({{x}})*G({{x}})=\int_{-\infty}^\infty L({{x}}')G({{x}}-{{x}}')\mathrm{d}{{x}}'.
\end{equation}
and equivalently
\begin{equation}
S_\mathrm{V}(x)=\sum_{n=({1-N})/{2}}^{(N-1)/2}V(x,x_0-n\Delta x,\Gamma,\sigma,A).
\end{equation}
We find for $\mathcal{F}_\mathrm{V}=\mathcal{F}[S_\mathrm{V}({{x}})]$ (\textit{c.f.}~Appendix~\ref{voigtder})
\begin{equation}
\label{voigtFT}
\mathcal{F}_\mathrm{V}=\mathrm{exp}\left[-\frac{1}{2}(\Gamma+\sigma^2k)k\right]\times\mathcal{F}_\mathrm{D}
\end{equation}
and observe that the sensitivity $\mathcal{M}_i\propto N$ still holds. Due to the convolution theorem the envelope $E_\mathrm{V}(k)$ defining the amplitudes of the $\mathcal{M}_{i}$ is given by
\begin{equation}
E_\mathrm{V}(k)=E_\mathrm{L}(k) E_\mathrm{G}(k)=\mathrm{exp}\left[-\frac{1}{2}(\Gamma+\sigma^2k)k\right]
\end{equation}
which is only parametrized by the Lorentzian line width $\Gamma$ and the Gaussian $\sigma$.
For illustration, figure~\ref{fig3} shows this for $N=8$, in addition to the Lorentzian and Gaussian cases. The Voigt FWHM $f_\mathrm{V}$ can be approximated with an accuracy of 0.02\%~\cite{olivero1977} by
\begin{equation}
\label{vfwhm}
f_\mathrm{V}=0.5346f_\mathrm{L}+\sqrt{0.2166f_\mathrm{L}^2+f_\mathrm{G}^2}
\end{equation}
where $f_\mathrm{L}$ denotes the Lorentzian FWHM $(f_\mathrm{L}=\Gamma)$.\\
Given a finite-periodic signal $S({{x}})$ consisting of $N$ Voigt-profile peaks $S_i(x)$, spaced by $\Delta{{x}}$, the line width contributions $\Gamma,\sigma$ of the individual signal peaks $S_i(x)$ can thus be determined by fitting $E_\mathrm{V}(k)$ to the $\mathcal{M}_{i}$, allowing to distinguish the Lorentzian contribution to the line width from that due to the  Gaussian. From equation~\ref{voigtFT} and figure~\ref{fig3} the behavior of $\mathcal{F}_\mathrm{V}$ is dominated by $E_\mathrm{L}(k)$ for small $k$ and by $E_\mathrm{G}(k)$ for large $k$, which allows for a quick qualitative analysis of the Lorentzian contribution by examination of the first few $\mathcal{M}_{i}$ - even by eye. In principle this applies also to the strong suppression or the slow decay of the tail by $E_\mathrm{G}(k)$ and $E_\mathrm{L}(k)$ respectively, but might not be unambiguously determined due to noise.

\subsection{Sensitivity and generalization of the method}
The Fourier transform simplifies the convolution integral to a product and thus, the number of profiles can be understood as the number of terms of a discrete Fourier transform leading to an increase of the sensitivity, linearly in $N$ (see equation~\ref{discreteFT} in the Appendix). This point of view can be taken in analogy to the $N-$periodic diffraction grating the $\mathcal{M}_{i}$ become more expressed and sharpen as $N$ increases with the limit of becoming the Dirac comb for $N\rightarrow\infty$. This allows us to extract values for  $\mathcal{M}_{i}$ even if $\mathcal{L}_{j}$ is below the noise level in the experimental setup. Thus, our result is particularly interesting for experimental applications where $N$ is large, although the case for small $N$ or even $N=1$ (\textit{c.f.}~reference~\cite{shin2014}) is possible, but not recommended.\\
Thanks to the simplification of the convolution integral and the linear increase in sensitivity with $N$, our method applied to other classes of line-profiles \cite{ruland1968} which are convolutions of two or more functions, should be as beneficial as for the Voigt profile. However, to which accuracy would have to be analyzed for each case separately.

\section{Example}
\label{application}
In this section we demonstrate the application of the proposed method to a multiplet signal, which in this case is a typical absorbance spectrum measured in transmission for a LiY$_{1-x}$Ho$_x$F$_4$ single crystal with $x=0.25\%$ at a temperature $T$ of $3.8$~K with a Fourier transform infrared (FTIR) spectrometer. The Bruker IFS125 spectrometer is based at the infrared beamline of the Swiss Light Source at Paul Scherrer Institut in Villigen, Switzerland \cite{albert2011b}. The spectrum in figure~\ref{fig4}(a) shows the hyperfine-split ground state to second excited state transition at $\sim23.3$~cm$^{-1}$ (700~GHz) in spectroscopic units [cm$^{-1}$]. Thanks to the ultra-high resolution FTIR in combination with the highly collimated, high-brillance infrared beam we achieved 0.001~cm$^{-1}$ (30~MHz) resolution. We refer to Matmon \textit{et al.}~\cite{matmon2016} for details on the general experimental setup and FTIR technique as well as for more extensive, spectroscopic work on LiY$_{1-x}$Ho$_x$F$_4$.

\subsection{Procedure}
\label{procedure}
Equation~\ref{voigtFT} leads to an algorithmic procedure for extracting the shape- and line width of a finite-periodic signal:
\begin{enumerate}
\item Calculate the Fourier transform of the finite-periodic signal and take the modulus.
\item Determine the maxima $\mathcal{M}_{i}$ defining the envelope $E_\mathrm{V}(k)$, which occur with periodicity $T_s$.
\item Fit the envelope function
\begin{equation}
E_\mathrm{V}(k,\Gamma,\sigma,P,c)=P\times\mathrm{exp}\left[{-\frac{1}{2}(\Gamma+\sigma^2k)k}\right]+c
\end{equation}
to the $\mathcal{M}_{i}$.
\end{enumerate}
We replace the analytical prefactor $|\mathcal{F}_\mathrm{D}|$ of the envelope with $P$ as the global scaling factor, combining factors from  the normalization of the Fourier transform, the single signal line area $A$ and experimental sensitivities. The offset constant $c$ accounts for nonzero white noise in the experiment. We note that if $\Delta x$ is known, the $\mathcal{M}_{i}$ might be more precisely determined thanks to $T_s=2\pi/\Delta x$. This simple procedure allows for algorithmic implementation.

\subsection{Application to experimental data}
\label{demonstration}
We apply the above procedure to the transmission spectrum in figure~\ref{fig4}(a). More details on the numerical discrete Fourier transform (DFT) procedure are found in the Appendix~\ref{ftdets}. Figure~\ref{fig4}(b) shows the modulus of the DFT coefficients $c_k$ of the unapodized (blue) and apodized (red) spectrum (details of apodization are in Appendix~\ref{ftdets}) in figure~\ref{fig4}(a) for $k<600$. The characteristic pattern of equation~\ref{voigtFT} is evident in figure~\ref{fig4}. The inset displays all Fourier coefficients. We observe reduced noise frequencies in $k$-space in figure~\ref{fig4}(b) for the apodized spectrum and a different global scaling factor $P$ with respect to the unapodized spectrum. However, the decay constants $\Gamma,\sigma$ are unchanged.\\
Figure~\ref{envfit} shows the DFT of the apodized spectrum and the $\mathcal{M}_{i}$. We apply a simple algorithm detecting the evenly spaced maxima $\mathcal{M}_{i}$ with period $T_s\sim45$ Fourier coefficients. We set a conservative threshold at $k=520$ to ensure a fiducial selection of the $\mathcal{M}_{i}$. We drop $\mathcal{M}_{k,0}$ as it is strongly affected by the Fourier transforms zero center peak. The result of the envelope $E(\Gamma,\sigma,P)$ fit for a Lorentzian, Gaussian and Voigt lineshape to the $\mathcal{M}_{i}$ is displayed in figure~\ref{envfit}. We observe that the Voigt profile yields the best results with $\Gamma=1.1\times10^{-2}\pm0.12\times10^{-3}$~cm$^{-1}$ and $\sigma=3.5\times10^{-3}\pm0.6\times10^{-3}$~cm$^{-1}$. With equation~\ref{fwhmg} we find $f_\mathrm{G}=8.5\times10^{-3}\pm1.4\times10^{-3}$~cm$^{-1}$ and with equation~\ref{vfwhm} $f_\mathrm{V}=15.7\times10^{-3}\pm1.6\times10^{-3}$~cm$^{-1}$. The same values within the fit precision are obtain by a direct Voigt profile fit to the individual signal peaks. The method determines that $f_\mathrm{L}$, the homogeneous part of the line width, is the slightly larger contribution.

\begin{figure}[htbp]
\centering\includegraphics[width=8.56cm]{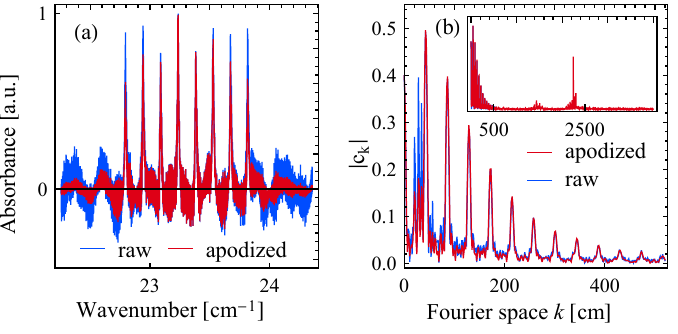}
\caption{\noindent (a) Unapodized and apodized absorbance spectrum of LiY$_{1-x}$Ho$_x$F$_4$~($x=0.3\%$) at a temperature T = 3.8~K, exhibiting an eightfold repeated signal due to the I = 7/2 nuclear hyperfine levels of the Ho atoms. The BH4T apodization function is centered to the spectrum and decays to zero on its ends. (b) DFT coefficient $|c_k|$ of data shown in (a). Inset: The characteristic pattern decays within the first 600 Fourier components. Around $k=1440$ and $k=2240$ strong noise contributions from Fabry-Perot interference on reflecting optical components are visible.}
\label{fig4}
\end{figure}
\begin{figure}[htbp]
\centering\includegraphics[width=5.6cm]{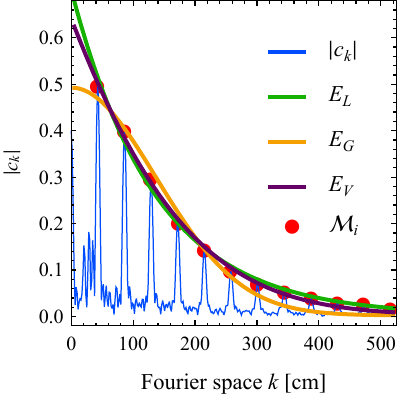}
\caption{\noindent Modulus of Fourier coefficients of apodized spectrum in figure~\ref{fig4} (a) (blue). Red points denote the selected $\mathcal{M}_{i}$. Solid lines represent the envelope fit.}
\label{envfit}
\end{figure}

\section{Robustness, noise stability and run time}
\label{robust}
We discuss the robustness of the method under deviations to the initial assumption of regular spacing and compare its performance to a direct least-squares fit method (DFM).

\subsection{Robustness}We examine the sensitivity of the fitting method to deviations from our initial assumption of finite-periodicity. In the case of unequal areas $A_i$, the individual area $A_i$ of the signal lines appear as an overall scaling factor which is already considered in section~\ref{application}. This allows the application of the method in cases where the individual signal line varies strongly in intensity \textit{e.g.} for rotational/vibrational spectra \cite{albert2011}. The extracted line width in case of differing individual line widths $\Gamma_i,\sigma_i$ is a weighted average.\\
As regular spacing is the underlying assumption, a variation of the line spacing $\Delta{{x}}$ directly modifies the $\mathcal{M}_{i}$ and thus the observed envelope $E(\Gamma,\sigma,P)$. Because of the line spacing variation the $N$ frequencies exhibit phase differences resulting in destructive interference and less expressed $\mathcal{M}_{i}$ (\textit{c.f.}~Appendix~\ref{detder}). This leads to an overestimation of the width and deviation from actual profile shape. We note that a second order hyperfine interaction in the example spectrum (for details see \cite{matmon2016}) results in a 2.7~\% deviation from equidistancy of the eight hyperfine lines. Nevertheless we obtain equivalent results for line width and shape, showing the robustness of the method against small (on the scale of $\lesssim\Delta{{x}}/N$) relaxation in finite-periodicity of $\lesssim3\%$.
Furthermore, if the variations $\Delta{{x}}_i$ are known, then $E(\Gamma,\sigma,P)$ could be adjusted to depend on the $\Delta{{x}}_i$.

\subsection{Noise-stability}
The noise spectrum is explicitly revealed with the FLA method. In case of a few isolated frequencies, as in \textit{e.g.} the inset in figure~\ref{fig4}, the noise does not affect the envelope fit in $k$-space and the method naturally implements a low-pass frequency filter. A particular strength of the method is that if there are a few frequencies between the $\mathcal{M}_{i}$ where there are what appear to be noise peaks, as is the case in figure~\ref{fig4} at \textit{e.g.} $k=70$. Such noise peaks will affect the DFM results and low-pass filters can not be applied. Consideration only of the primary peaks at $\mathcal{M}_{i}$ provides an impartial filter of the associated noise\\
Assuming Gaussian RMS noise for artificial spectra with up to 50\% of the signal amplitude, we find the FLA method to be comparable to DFM within the statistical parameter errors (\textit{c.f.}~Appendix~\ref{rms}). This holds true for different ratios of $\Gamma/\sigma\in{0.1,1,10}$. We find that the precision for both, FLA and DFM strongly decreases for $\Gamma/\sigma<0.1$ and $\Gamma/\sigma>10$. We observe a systematic error for the FLA method in the form of an increase of the Gaussian line width contribution with increasing RMS noise amplitude. We attribute this to the fact that our analysis in Fourier space reflects the noise spectrum convolved with the signal and thus adds to the Gaussian line width contribution. Although the DFM method may occasionally exhibit a more precise performance, the FLA method extracts line width and shape information algorithmically, with less computational power and fewer data points, as detailed in the next subsection.

\subsection{Run time}
We note the significantly different computational intensity of both methods. On our system (Intel(R) Xeon(R)~E5-2670~0~@2.6~GHz, 2 processors with 192~GB usable RAM) DFM is at least one to two orders of magnitude slower than the FLA method. A test on 1000 data sets with $N=8$ peaks, 1000 data points, and varying Gaussian RMS noise yielded 1.6~s run time for the FLA method and 2543~s for DFM. Runtimes are dataset specific and further (algorithm) optimization may reduce the run time discrepancy. For fundamental reasons, however, a run time difference will persist as the FLA method allows to reduce the parameter space to four $(\Gamma,\sigma,P,c)$, whereas a direct fit method has to consider at least six ($\Gamma,\sigma$,area $A$, position $x_0$, spacing $s$, offset $c$) under the assumption of identical areas $A_i$, which is rarely fulfilled. Any individual consideration of a parameter in the direct fit scales the parameter space with $N$. This increases the run time for the DFM method significantly. Furthermore, the final envelope fit is performed on a fraction of the initial data points. In case of insufficient precision of the FLA method, it may well serve as a fast pre-analysis for good DFM start parameters.

\section{Conclusion}
\label{conclusion}
We have introduced and demonstrated a method of extracting the line-width and shape of a multiplet signal. The basic idea and mathematics are actually straightforward generalizations of the Debye-Waller effect \cite{alsnielsen2011} in scattering techniques, where the intensities of Bragg peaks shrink with increasing order in proportion to an envelope function decaying as a Gaussian whose width in reciprocal space is inversely proportional to the uncertainty in the positions of the atoms in real space. We offer a direct and computationally efficient way of quantifying the Lorentzian and Gaussian parts of multiple Voigt profiles by harnessing the regular spacing of the signal. The method benefits from a linear increase in sensitivity with the number of profiles of the multiplet. Furthermore, we highlighted that the FLA method is comparable to a direct least square fit, but is less computationally intensive and exhibits significant advantages in the presence of low-frequency noise. Finally, the FLA method might serve as a fast and algorithmic pre-analysis for starting parameter determination cases where the assumption of periodic repetition of the same lineshape is broken. Potential applications are ample. X-ray scattering may benefit where quasiperiodic structures are sampled within finite real space windows for problems such as integrated circuit microscopy \cite{holler2017,holler2019}. Our method could also be applied to optical comb-based spectroscopies \cite{haensch2005} for precision measurements in atomic and solid-state physics.
%%%%%%%%%%%%%%%%%%%%%%%%%%  Appendix  %%%%%%%%%%%%%%%%%%%%%%%%%%
\section*{Appendix}
\label{suppl}
\renewcommand{\thesubsection}{\Alph{subsection}}
\subsection{Definitions}
\label{defs}
The Fourier transform we use throughout the article is defined as
\begin{equation}
\mathcal{F}[S({{x}})]=1/\sqrt{2\pi}\int_{-\infty}^\infty S({{x}})\mathrm{exp}[- i k {{x}}]\mathrm{d}{{x}}
\end{equation}
We use the following definition of a Dirac comb $\Sha$:
\begin{equation}
\Sha_{\Delta x}(x)=\frac{1}{\Delta x}\Sha\left(\frac{x}{\Delta x}\right)=\sum_{n=-\infty}^\infty\delta(x-n\Delta x)
\end{equation}
where $\delta(x)$ denotes the Dirac delta function. The Fourier series of the Dirac comb is given by
\begin{equation}
\label{dcfour}
\begin{split}
\Sha_{\Delta x}(x)=\frac{1}{\sqrt{2\pi}}\sum_{n=-\infty}^{\infty}\mathrm{exp}({-ikn\Delta x})
=\frac{1}{\sqrt{2\pi}}\left[1+2\sum_{n=1}^{\infty}\mathrm{cos}({kn\Delta x})\right]
\end{split}
\end{equation}
which directly yields another interpretation of equation~\ref{diracFT}: It is the sum of the first $N$ terms of the Fourier series of the Dirac comb $\Sha_{\Delta x}(x)$ for odd $N$.\\
The rectangular window function of height $h$ and width $w$ is defined as
\begin{equation}
\mathrm{rect}(h,w)=h\times\mathrm{rect}\left(\frac{x}{w}\right)
\end{equation}
where we use the standard definition for the unit rectangular function $\mathrm{rect}(x)=1~\forall x \in \{-1/2,1/2\}$ and 0 otherwise.\\
The model where the individual signal peaks $S_i(x)=G({{x}})$ are Gaussian (\textit{c.f.} figure~\ref{fig2}(a) and (b)~) is given by
\begin{equation}
\label{gauss}
S_\mathrm{G}({{x}})=\frac{A}{\sqrt{2\pi}\sigma}\sum_{n=({1-N})/{2}}^{(N-1)/2}\mathrm{exp}\left[-\frac{({{x}}-x_0-n\Delta{{x}})^2}{2\sigma^2}\right]
\end{equation}
with standard deviation $\sigma$, center signal line position $x_0$, spacing $\Delta{{x}}$ and integrated area $A$. The Fourier transform of $S_\mathrm{G}({{x}})$ (with $x_0=0$) is given by
\begin{equation}
\label{gaussFTapp}
\mathcal{F}_\mathrm{G}=\mathcal{F}\left[S_\mathrm{G}({{x}})\right]=\mathrm{exp}\left(-\frac{1}{2}\sigma^2 k^2\right)\mathcal{F}_\mathrm{D}
\end{equation}

\subsection{Detailed derivation}
\label{detder}
The Dirac model $S_\mathrm{D}({{x}})$ is given by
\begin{equation}
\label{dirac}
S_\mathrm{D}({{x}})=A\sum_{n=({1-N})/{2}}^{(N-1)/2}\delta({{x}}-n\Delta{{x}})
\end{equation}
assuming symmetry around ${{x}}=0$. With $\{A,n,\Delta{{x}},k\}\in\mathbb{R}$ we obtain for $\mathcal{F}_\mathrm{D}$
\begin{equation}
\begin{split}
\mathcal{F}_\mathrm{D}=\frac{1}{\sqrt{2\pi}}\int_{-\infty}^\infty A\sum_{n=({1-N})/{2}}^{(N-1)/2}\delta({{x}}-n\Delta{{x}})\times\mathrm{exp}({-ikx})~\mathrm{d}x
=\frac{A}{\sqrt{2\pi}}\sum_{n=({1-N})/{2}}^{(N-1)/2}\mathrm{exp}({-ikn\Delta x})
\end{split}
\label{discreteFT}
\end{equation}
which, for $N$ even yields
\begin{equation}
\label{neven}
\begin{split}
\mathcal{F}_\mathrm{D}\overset{N~\mathrm{even}}{=}\frac{A}{\sqrt{2\pi}}\sum_{n=1/2}^{(N-1)/2}\mathrm{exp}({-ikn\Delta x})+\mathrm{exp}({ikn\Delta x})
=A\sqrt{\frac{2}{\pi}}\sum_{n=1/2}^{(N-1)/2}\mathrm{cos}(n\Delta x k)
\end{split}
\end{equation}
and for odd $N$
\begin{equation}
\label{nodd}
\begin{split}
\mathcal{F}_\mathrm{D}\overset{N~\mathrm{odd}}{=}\frac{A}{\sqrt{2\pi}}\sum_{n=0}^{(N-1)/2}\mathrm{exp}({-ikn\Delta x})+\mathrm{exp}({ikn\Delta x})
=A\sqrt{\frac{2}{\pi}}\left[\frac{1}{2}+\sum_{n=0}^{(N-1)/2}\mathrm{cos}(n\Delta x k)\right]
\end{split}
\end{equation}
For even $N$ we further use
\begin{equation}
\label{sumsindet}
\begin{split}
&\sum_{n=1/2}^{(N-1)/2}\mathrm{cos}(n\Delta x k)=\sum_{n=0}^{N/2-1}\mathrm{cos}\left[\left(n+\frac{1}{2}\right)\Delta x k\right]
=\sum_{n=0}^{N/2-1}\mathrm{Re}\left\{\mathrm{exp}\left[i\left(n+\frac{1}{2}\right)\Delta x k\right]\right\}\\
&=\mathrm{Re}\left[\mathrm{exp}\left(\frac{i\Delta x k}{2}\right)\sum_{n=0}^{N/2-1}\mathrm{exp}\left(in\Delta x k\right)\right]
=\mathrm{Re}\left[\mathrm{exp}\left(\frac{i\Delta x k}{2}\right)\frac{1-\mathrm{exp}(i N \Delta x k)}{1-\mathrm{exp}(i\Delta x k)}\right]\\
&=\mathrm{Re}\left[\mathrm{exp}\left(\frac{i N \Delta x k }{4}\right)\frac{\mathrm{exp}\left(-\frac{i N \Delta x k }{4}\right)-\mathrm{exp}\left(\frac{i N \Delta x k }{4}\right)}{\mathrm{exp}\left(-\frac{i\Delta x k}{2}\right)-\mathrm{exp}\left(\frac{i\Delta x k}{2}\right)}\right]\\
&=\frac{\mathrm{cos}(N \Delta x k/4)\mathrm{sin}(N \Delta x k/4)}{\mathrm{sin}(\Delta x k/2 )}=\frac{\mathrm{sin}(N\Delta x k/2)}{\mathrm{sin}(\Delta x k/2)}
\end{split}
\end{equation}
and we leave the case for the odd $N$ to the reader. From equation~\ref{neven} we conclude that
\begin{equation}
\label{finderiv}
\begin{split}
\mathcal{F}_\mathrm{D}=\frac{A}{\sqrt{2\pi}}\frac{\mathrm{sin}(N\Delta x k/2)}{\mathrm{sin}(\Delta x k/2)}~\forall N\ge2\in\mathbb{N}.
\end{split}
\end{equation}
Equation~\ref{finderiv} shows that the number of signal lines $N$ of $S({{x}})$ determines the amplitude of the $\mathcal{M}_{i}= N A/\sqrt{2\pi}~\forall i$. This is determined with the rule of Bernoulli-de l'H\^{o}spital where numerator $[\mathrm{sin}(N\Delta x k/2)]$ and denominator $[\mathrm{sin}(\Delta x k/2)]$ vanish which is the case for $\Delta{{x}k/2}~\mathrm{mod}~\pi=0$. Furthermore, $\mathcal{L}_\mathrm{j}\sim A/\sqrt{2\pi}~\forall j$ as $\mathcal{L}_{j,\mathrm{min}}$ denotes the point where numerator and denominator are equal to one. Note that for even $N$ this condition is never exactly met and $\mathcal{L}_{j,\mathrm{min}}$ slightly smaller than 1 but still independent of $N$. These results lead to
\begin{equation}
\label{snrapp}
\frac{\mathcal{M}_i}{\mathcal{L}_{j,\mathrm{min}}}\propto N
\end{equation}
which is the scaling of the sensitivity of the FLA method with the number $N$ of individual signal profiles $S_i(x)$, in analogy to a finite diffraction grating. This result can be alternatively understood as the approximate ratio of the main maximum to the first side lobe of a sinc-function, as shown in the next subsection.

\subsection{Alternative derivation}
\label{altder}
A sum of $N$ delta functions centered around 0 and spaced by $\Delta x$ can be written as
\begin{equation}
\begin{split}
S_\mathrm{D}({{x}})=&A\sum_{n=({1-N})/{2}}^{(N-1)/2}\delta({{x}}-n\Delta{{x}})~\forall N\in\mathbb{N}\ge2
\end{split}
\end{equation}
or as
\begin{equation}
S_\mathrm{D}({{x}})=A\Sha_{\Delta x}(x) \mathrm{rect}\left(\frac{x}{N\Delta x}\right)~\forall N\in\mathbb{N}\ge2
\end{equation}
For $\mathcal{F}[S_\mathrm{D}({{x}})]\equiv\mathcal{F}_\mathrm{D}$ follows with the application of the convolution theorem:
\begin{equation}
\label{sincsum}
\begin{split}
\mathcal{F}_\mathrm{D}&=A\mathcal{F}\left[\Sha_{\Delta x}(x) \mathrm{rect}\left(\frac{x}{N\Delta x}\right)\right]
=A\mathcal{F}\left[\Sha_{\Delta x}(x)\right]*\mathcal{F}\left[ \mathrm{rect}\left(\frac{x}{N\Delta x}\right)\right]\\
&=\frac{A}{\Delta x}\Sha_{\frac{1}{\Delta x}}(k)*\frac{N\Delta x}{\sqrt{2\pi}}\mathrm{sinc}\left(\frac{N\Delta x}{2}k\right)
=\frac{A N}{\sqrt{2\pi}}\sum_{m=-\infty}^{\infty}\mathrm{sinc}\left[\frac{N\Delta x}{2}\left(k-\frac{m}{\Delta x}\right)\right]
\end{split}
\end{equation}
which reveals equation~\ref{diracFT} to be an infinite sum of sinc functions.

\subsection{Detailed derivation of Voigt model}
\label{voigtder}
We present the detailed derivation for $\mathcal{F}_\mathrm{V}$.
\begin{equation}
\label{voigtcase}
\begin{split}
\mathcal{F}_\mathrm{V}&=\mathcal{F}\left[\sum_{n=({1-N})/{2}}^{(N-1)/2}V(x,x_0-n\Delta x,\Gamma,\sigma,A)\right]
=\sum_{n=({1-N})/{2}}^{(N-1)/2}\mathcal{F}\left[V(x,x_0-n\Delta x,\Gamma,\sigma,A)\right]\\
&=\sum_{n=({1-N})/{2}}^{(N-1)/2}\mathcal{F}\left[L(x,x_0-n\Delta x,\Gamma,A)*G(x,0,\sigma,1)\right]\\
&=\sum_{n=({1-N})/{2}}^{(N-1)/2}\mathcal{F}\left[L(x,x_0-n\Delta x,\Gamma,A)\right]\times\mathcal{F}\left[G(x,0,\sigma,1)\right]
\end{split}
\end{equation}
which for $x_0=0$ simplifies to
\begin{equation}
\label{voigtcaseB}
\begin{split}
\mathcal{F}_\mathrm{V}&=\frac{A}{\sqrt{2\pi}}\sum_{n=({1-N})/{2}}^{(N-1)/2} \mathrm{exp}\left({ i k n\Delta x-\frac{1}{2}\Gamma}\right)\times \mathrm{exp}\left({-\frac{1}{2}\sigma^2k^2}\right)\\
&=\frac{A}{\sqrt{2\pi}}\sum_{n=({1-N})/{2}}^{(N-1)/2} \mathrm{exp}\left[{ i k n\Delta x-\frac{1}{2}(\Gamma+k\sigma^2)k}\right]\\
&=\frac{A}{\sqrt{2\pi}}\mathrm{exp}\left[{-\frac{1}{2}(\Gamma+k\sigma^2)k}\right]\sum_{n=({1-N})/{2}}^{(N-1)/2} \mathrm{exp}({i k n\Delta x})\\
&\overset{N~\mathrm{even}}{=}A\sqrt{\frac{2}{\pi}}\mathrm{exp}\left[{-\frac{1}{2}(\Gamma+k\sigma^2)k}\right]\sum_{n=1/{2}}^{(N-1)/2} \mathrm{cos}(k n\Delta x )
\end{split}
\end{equation}
We leave the case of odd $N$ to the reader.

\subsection{Details on numerical Fourier transform and apodization}
\label{ftdets}
Prior to the Fourier transform, the application of an apodization function to the spectrum is recommended. High frequency noise and distortion effects of lineshapes due to the DFT on finite-sized signals are minimized. The red trace in figure~\ref{fig4}(a) is apodized by a 4~\wn wide Blackman-Harris 4-term (BH4T) apodization function \cite{harris1978,nuttall1981}, centered at 23.3~\wn. The unit width BH4T$(x)$ function $-1/2\le x\le1/2$ is defined as
\begin{equation}
\mathrm{BH4T}(x)=0.035875+0.48829\times\mathrm{cos}(2\pi x)+0.14128\times\mathrm{cos}(4\pi x)+0.0.01168\times\mathrm{cos}(6\pi x)
\end{equation}
and $\mathrm{BH4T}(x)=0~\forall~|x|>1/2$ \cite{nuttall1981}.

\subsection{RMS noise test}
\label{rms}
\begin{figure}[htbp]
\centering\includegraphics[width=8.56cm]{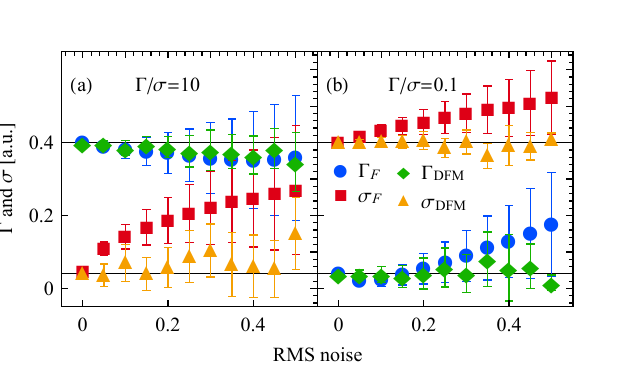}
\caption{\noindent RMS noise test. (a) Extracted fit results $\Gamma_F,\sigma_F$ for \mbox{$\Gamma/\sigma=10$} under RMS noise with the FLA method and direct least-squares fit ($\Gamma_\mathrm{DFM},\sigma_\mathrm{DFM}$). Gridlines denote the initial values before RMS noise application (b) Same as (a) with inverse contribution ratio $\Gamma/\sigma=0.1$}
\label{figapp1}
\end{figure}

For the RMS noise test we create artificial data, where we add different amplitudes of RMS noise and apply the FLA as well as the DFM method with a Voigt lineshape model, for the case of $N=8$ peaks. The results are shown in figure~\ref{figapp1}. Gridlines denote the initial values before RMS noise application. For contribution ratios $\Gamma/\sigma=0.1$ the FLA method tends to overestimate the overall line width. We attribute this to a systematic error, which pushes the ratio in the limit where the methods precision strongly decreases.

\newpage
\section*{Funding}
\label{funding}
Swiss National Science Foundation, Grant No. 200021\_166271.

\section*{Acknowledgments}
\label{ack}
FTIR spectroscopy data were collected at the X01DC beamline of the Swiss Light Source, Paul Scherrer Institut, Villigen, Switzerland. We are grateful to G.~Matmon, M.~Grimm and J.W.~Spaak for helpful discussions and G.~Matmon and S.~Gerber for critically reviewing the manuscript.

\section*{Disclosures}
The authors declare that they have no competing interests.

%%%%%%%%%% If using BibTeX:
\bibliography{References}

\end{document}